# Observer Design for Takagi-Sugeno Descriptor System with Lipschitz Constraints

Kilani Ilhem[1], Jabri Dalel[2], Bel Hadj Ali Saloua[1] and Abdelkrim[1]

Mohamed Naceur

[1]MACS Unit - National Engineering School of Gabès Tunisia, University of Gabès
kilani.ilhem@gmail.com
[2]University of Reims Champagne-Ardenne CReSTIC EA3804, Moulin de la Housse
BP 1039, 51687 Reims Cedex 2, France.
dalel.jabri@yahoo.fr

**ABSTRACT**

*This paper investigates the design problem of observers for nonlinear descriptor systems described by Takagi-Sugeno (TS) system; Depending on the available knowledge on the premise variables two cases are considered. First a TS descriptor system with measurables premises variables are proposed. Second, an observer design which satisfying the Lipschitz condition is proposed when the premises variables are unmeasurables. The convergence of the state estimation error is studied using the Lyapunov theory and the stability conditions are given in terms of Linear Matrix Inequalities (LMIs). Examples are included to illustrate those methods.*

## 1. INTRODUCTION

Since more than two decades ago, Takagi–Sugeno (T–S) fuzzy models have attracted wide attention from scientists and engineers, essentially because the well-known fuzzy models can effectively approximate a wide class of nonlinear systems. Relaxed sufficient conditions for fuzzy controllers and fuzzy observers are proposed in via a multiple Lyapunov function approach [1].

System modeling by descriptor method has an important use in the literature since it represents many class of non linear system. The enhancement of the modeling ability is due to the structure of the dynamic equation which encompasses not only dynamic equations, but also algebraic relations [2].Since both TS and descriptor formalism are attractive in the field of modeling, the TS representation has been generalized to descriptor systems. The stability and the design of state-feedback controllers for T-S descriptors systems are characterized via LMI in [3], [4].

System modeling by descriptor method has an important use in the literature since it represents many class of non linear system.

In other hand, many problems in decision making, control and monitoring require a state estimation based on a dynamic system model. A generic method for the observer design, valid for all types of nonlinear systems, has not yet been developed. This problem received considerable attention in the last three decades; it is of great importance in theory and practice, since there are many situations where premise variable are inaccessible.

In this paper, we propose to design a state observer to estimate exactly the descriptor's states. For a TS fuzzy model, well-established methods and algorithms can be used to design observers that estimate measurable states. Several types of such observers have been developed for TS fuzzy systems, among which: fuzzy Thau–Luenberger observers reduced-order observers and sliding-mode observers. In general, the design methods for observers also lead to an LMI feasibility problem [5], [6], [7], [8].

For the unmeasurables premises variables many searchers studied the observer design for non linear descriptor system with lipschitz constraint [9] [10][11][12].The fuzzy TS systems are studied in [13]. The main objective of this paper is to develop an observer design for TS descriptor system with lipschitz constraint. Under some sufficient conditions, the design of the observer is reduced to the determination of a matrix. The choice of this parameter is performed by solving strict LMIs (linear matrix inequalities).

The outline of the paper is as follows. First, the class of studied systems and observer are given. Second problem formulation for observer with measurable variables is dealt. Third the Lipschitz observer will be studied. Finally we present an example to illustrate the effectiveness of the proposed method. Concluding remarks finish this paper.

## 2. SYSTEM DESCRIPTION AND PRELIMINARIES

### 2.1. Descriptor System

Let us consider the class of non linear descriptor system which is defined as:

$$E(x(t))\dot{x}(t) = A(x(t))x(t) + B(x(t))u(t)$$
$$y(t) = C(x(t))x(t) \qquad (1)$$

Where $x(t) \in R^n$, $u(t) \in R^q$ and $y(t) \in R^m$ represent respectively the state, the control input and the output vectors; $A(x(t)), E(x(t)) \in R^{n \times n}$, $B(x(t)) \in R^{n \times m}$ and $C(x(t)) \in R^{q \times n}$ are non linear matrices functions. For simplicity, we should always consider that $E(x(t))$ is regular for each $x(t) \in R^n$.

Then the TS descriptor system is given as:

$$\sum_{k=1}^{l} v_k(z(t)) E_k \dot{x}(t) = \sum_{i=1}^{r} h_i(z(t))(A_i x(t) + B_i u(t))$$
$$y(t) = Cx(t) \qquad (2)$$

Where $z$ represent the premise variable; $h_i(z(t))$ and $v_k(z(t))$ represent respectively the right and the left activating function; $r$ and $l$ is respectively the right and the left number of fuzzy rules ; $E_k$ $A_i$, $B_i$ and $C$ are constant matrices;

Where

$$h_i(z(t)) = \frac{w_i(z(t))}{\sum_{i=1}^{r} h_i(z(t))} \quad , \quad w_i(z(t)) = \prod_{i=1}^{g} F_{ij}(z_j(t)) \qquad (3)$$

$h_i(z(t))$ has also the same characteristic as $v_k(z(t))$ and it satisfies:

$$h_i(z(t)) = 1 \text{ and } h_i(z) > 0 \text{ for } i = 1,...,r \qquad (4)$$

The passage from the nonlinear system to the TS descriptor system is obtained by the sector nonlinearity approach. In this case, the TS descriptor matches exactly the nonlinear model in a compact set of the state variables .

### 2.2. Descriptor observer

In general a TS observer is defined by the interconnection between many locals Luenberger observers [14], [15]. It is written as:

$$\dot{\hat{x}}(t) = \sum_{i=1}^{r} h_j(\hat{z}(t))\left[A_i\hat{x}(t) + B_iu(t)L_i(y(t) - \hat{y}(t))\right] \qquad (5)$$
$$\hat{y}(t) = C\hat{x}(t)$$

Where $\hat{x}(t)$ and $\hat{z}(t)$ represent respectively the estimate state and the estimate premise variable; $L_i$ are the observer gain matrices.

In this study, the proposed observer is in the descriptor form. Using the general TS descriptor form, it is possible to suppose a nonlinear observer based on this form:

$$\sum_{k=1}^{l} v_k(z(t))E_k\dot{\hat{x}}(t) = \sum_{j=1}^{r} h_j(\hat{z}(t))\left[A_i\hat{x}(t) + B_iu(t)L_i(y(t) - \hat{y}(t))\right]$$
$$\hat{y}(t) = C\ \hat{x}(t) \qquad (6)$$

There are two cases to define witch related to the accessibility of the premises variables. In the first time we assume that the variables $z(t)$ are real time available $\hat{z}(t) = z(t)$ and thus so are the weighting functions $h_i(\hat{z}(t)) = h_i(z(t))$ and $v_k(\hat{z}(t)) = v_k(z(t))$. But, in many practical situations, these premise variables depend on the state variables that are not always accessible. Then two cases are considered.

- measurables premises variables
- unmeasurables premises variables.

## 3. OBSERVER DESIGN

### 3.1. Measurables Premises Variables

This section is devoted to the state estimation. In fact a descriptor form of observer will be considered. The following theorem presents the main result.

Theorem1: The convergence of the error estimation between the system (2) and the observer (6) is ensured if there exist: $P_1 = P_1^T > 0$ and $P_3, L_i, Y_{3i}$ for $i = 1,...,r$ and $k = 1,...,l$ such that:

$$\begin{bmatrix} A_i^T P_3 - C^T Y_{3i}^T + P_3^T A_i - Y_{3i}C & (*) \\ P_1 - E_k^T P_3 + P_3^T A_i - Y_{3i}C & -E_k^T P_3 - P_3^T E_k \end{bmatrix} < 0 \qquad (7)$$

As usual, a star (*) indicates a transpose quantity in a symmetric matrix.

Proof:
Away to obtain sufficient convergence conditions for descriptor fuzzy observer is to consider the error estimation as:
$$e(t) = x(t) - \hat{x}(t) \qquad (8)$$

Its time derivative is given by:
$$\dot{e}(t) = \dot{x}(t) - \dot{\hat{x}}(t) \qquad (9)$$

According to equation (2) equation (6) becomes:
$$\sum_{k=1}^{l} v_k(z(t))E_k\dot{\hat{x}}(t) = \sum_{i=1}^{r} h_i(z(t))\left((A_i\hat{x}(t) + B_iu(t)) + L_i(Cx(t) - C\hat{x}(t))\right) \qquad (10)$$

The difference between this last equations and equation (2) is given by:

$$\sum_{k=1}^{l} v_k(z(t))E_k\dot{e}(t) = \sum_{i=1}^{r} h_i(z(t))(A_i - L_iC)e(t) \tag{11}$$

Let us consider the augmented error vector is given by:

$$\bar{e}(t) = \begin{bmatrix} e^T(t) & \dot{e}^T(t) \end{bmatrix}^T \tag{12}$$

Equations (11) can be written:

$$\bar{E}\dot{\bar{e}}(t) = \bar{A}_{ik}\bar{e}(t) \tag{13}$$

With

$$\bar{E} = \begin{bmatrix} I & 0 \\ 0 & 0 \end{bmatrix} \text{ and } \bar{A}_{ik} = \begin{bmatrix} 0 & I \\ A_i - L_iC & -E_k \end{bmatrix} \tag{14}$$

In order to find an asymptotic convergence error, we consider the Lyapunov candidate function $V(\bar{e}(t)) = \bar{e}(t)^T \bar{E}\bar{P}\bar{e}(t), P = P^T, P \in R^{n \times n}$ with:

$$\bar{E}P = P^T\bar{E} > 0, P = \begin{bmatrix} P_1 & P_2 \\ P_3 & P_4 \end{bmatrix}, P_1 = P_1^T > 0 \text{ and } P_2 = 0 \tag{15}$$

The negativity of the Lyapunov function is assumed by:

$$\dot{V}(\bar{e}(t)) = \dot{\bar{e}}(t)^T \bar{E}P\bar{e}(t) + \bar{e}(t)^T \underbrace{\bar{E}P}_{P^T\bar{E}}\dot{\bar{e}}(t) < 0 \tag{16}$$

In others words, obviously, with (14), the condition (16) becomes:

$$\bar{e}(t)^T \left(\bar{A}_{ik}^T P + P^T \bar{A}_{ik}\right)\bar{e}(t) < 0 \tag{17}$$

Then for $i = 1,...,r$ and for $k = 1,...,l$ we obtain:

$$\begin{bmatrix} 0 & (A_i - L_iC)^T \\ I & -E_k^T \end{bmatrix} * \begin{bmatrix} P_1 & 0 \\ P_3 & P_4 \end{bmatrix} + \begin{bmatrix} P_1^T & P_3^T \\ 0 & P_4^T \end{bmatrix} * \begin{bmatrix} 0 & I \\ A_i - L_iC & -E_k \end{bmatrix} < 0 \tag{18}$$

$$\begin{bmatrix} A_i^T P_3 - C^T L_i^T P_3 + P_3^T A_i - P_3^T L_i C & (*) \\ P_1 - E_k^T P_3 + P_4^T A_i - P_4^T L_i C & -E_k^T P_4 - P_4^T E_k \end{bmatrix} < 0 \tag{19}$$

In order to solve strict LMIs, we suppose that $P_4 = P_3$, then

$$\begin{bmatrix} A_i^T P_3 - C^T L_i^T P_3 + P_3^T A_i - P_3^T L_i C & (*) \\ P_1 - E_k^T P_3 + P_3^T A_i - P_3^T L_i C & -E_k^T P_3 - P_3^T E_k \end{bmatrix} < 0 \tag{20}$$

To have terms of Linear Matrix Inequalities we consider a bijectif changement $Y_{3i} = P_3^T L_i$:

$$\begin{bmatrix} A_i^T P_3 - C^T Y_{3i}^T + P_3^T A_i - Y_{3i} C & (*) \\ P_1 - E_k^T P_3 + P_3^T A_i - Y_{3i} C & -E_k^T P_3 - P_3^T E_k \end{bmatrix} < 0 \tag{21}$$

### 3.2 Unmeasurables Premises Variables

This part addresses the issues of observer design for a class of descriptor systems with Lipschitz constraint.The general form for the TS descriptor system is in the form of equation (2).To avoid to have a complex equation, we will suppose that all the $E_k = E$ for $k = 1,...,l$ .Then we have a new form of TS descriptor system given as below:

$$E\dot{x}(t) = \sum_{i=1}^{r} h_i(z(t))(A_i x(t) + B_i u(t))$$
$$y(t) = Cx(t)$$
(22)

Then the observer corresponds to this extension of TS descriptor system will be defined as:

$$E\dot{\hat{x}}(t) = \sum_{j=1}^{r} h_j(\hat{z}(t))[A_i \hat{x}(t) + B_i u(t) L_i(y(t) - \hat{y}(t))]$$
$$\hat{y}(t) = C\hat{x}(t)$$
(23)

In this section, the purpose is to suggest a method for the design of observer for TS descriptor system. The following hypothecs and lemma will be used in the development.

Hypothec 1: The activating function is lipschitz:
$$|h_i(x(t)) - h_i(\hat{x}(t))| \leq \gamma_i |x(t) - \hat{x}(t)|$$
(24)

$$|h_i(x(t))x(t) - h_i(\hat{x}(t))\hat{x}(t)| \leq m_i |x(t) - \hat{x}(t)|$$
(25)

Where $\gamma_i$ and $m_i$ positives scalars Lipschitz constants.

Hypothec 2: The control input u (t) is bounded:
$$\|u(t)\| \leq \beta_1, \beta_1 > 0$$
(26)

Lemma1
For all matrices X and Y, $\lambda$ a positive scalar, the following property is given as:
$$X^T Y + Y^T X \leq \lambda X^T X + \lambda^{-1} Y^T Y, \quad \lambda > 0$$
(27)

The following theorem shows the most important result.

Theorem2: The convergence of the error estimation between the system (22) and the observer (23) tend asymptotic to zero is ensured if there exist: $P \in R^{n \times n}$ and $Q \in R^{n \times n}$ symmetric positive defined, the matrix $K_{1i} \in R^{n \times q}$ and positives scalars $\lambda_1$, $\lambda_2$ such that:

$$A_0^T P + PA_0 - C^T K_i^T - K_i C < -Q$$
(28)

$$\begin{bmatrix} -Q + \lambda_1 m_i^2 I & P\bar{A}_i & P\bar{B}_i & n_i \gamma I \\ \bar{A}_i^T P & -\lambda_1 I & 0 & 0 \\ \bar{B}_i^T P & 0 & -\lambda_2 I & 0 \\ n_i \gamma I & 0 & 0 & -\lambda_2 I \end{bmatrix} < 0$$
(29)

$$\gamma - \beta_1 \lambda_2 \geq 0$$
(30)

$$EP = P^T E > 0$$
(31)

The gain of the observer is given by:
$$L_i = P^{-1} K_i$$
(32)

Proof:
The basic idea is to concept a descriptor observer which verifies Lipschitz condition by introducing some change in the system. Hence, the matrices $A_i, B_i$ and $C_i$ will be written as :

$$A_0 = \sum_{i=1}^{r} A_i, \quad \bar{A}_i = A_i - A_0 \tag{33}$$

$$B_0 = \sum_{i=1}^{r} B_i, \quad \bar{B}_i = B_i - B_0 \tag{34}$$

$$C_0 = \sum_{i=1}^{r} C_i, \quad \bar{C}_i = C_i - C_0 \tag{35}$$

The system (23) will be written as:

$$E\dot{x}(t) = A_0 x(t) + B_0 u(t) + \sum_{i=1}^{r} h_i(x(t))(\bar{A}_i x(t) + \bar{B}_i u(t))$$
$$y(t) = C\, x(t) \tag{36}$$

Then the observer equation will be changed as:

$$E\dot{\hat{x}}(t) = A_0 \hat{x}(t) + B_0 u(t) \sum_{i=1}^{r} h_i(\hat{x}(t))(\bar{A}_i \hat{x}(t) + \bar{B}_i u(t)) + L_i(y(t) - \hat{y}(t))$$
$$\hat{y}(t) = C\, \hat{x}(t) \tag{37}$$

The estimation error is given as:

$$e(t) = x(t) - \hat{x}(t) \tag{38}$$

Consequently, the augmented state estimation error obeys to the following nonlinear system:

$$E\dot{e}(t) = \sum_{i=1}^{r} h_i(\hat{x}(t))(A_0 - L_i C) e(t) + \sum_{i=1}^{r} \bar{A}_i \left( h_i(x(t)) x(t) - h_i(\hat{x}(t))\hat{x}(t) \right) + \bar{B}_i \left( h_i(x(t)) - h_i(\hat{x}(t)) \right) u(t) \tag{39}$$

To show the effeteness of hypothecs we suppose as below:

$$\begin{cases} \delta_i(t) = h_i(x(t)) x(t) - h_i(\hat{x}(t)) \hat{x}(t) \\ \Delta_i(t) = h_i(x(t)) - h_i(\hat{x}(t)) \\ \Phi_i(t) = A_0 - L_i C \end{cases} \tag{40}$$

Then the dynamic error will be defined as:

$$E\dot{e}(t) = \sum_{i=1}^{r} h_i(\hat{x}(t)) \Phi_i(t) e(t) + \sum_{i=1}^{r} \bar{A}_i \delta_i(t) + \bar{B}_i \Delta_i(t) u(t) \tag{41}$$

In order to find an asymptotic convergence error, we consider the quadratic Lyapunov function $V(e(t)) = e(t)^T E P e(t), P = P^T, P \in R^{n \times n}$ with: $EP = P^T E > 0$

Then the time derivative of the Lyapunov function will be written as:

$$\dot{V}(e(t)) = \dot{e}(t)^T E P e(t) + e(t)^T \underbrace{EP}_{P^T E} \dot{e}(t) \tag{42}$$

Then

$$\dot{V}(e(t)) = \sum_{i=1}^{r} \delta_i(t)^T \bar{A}_i^T P e(t) + e(t)^T P \bar{A}_i \delta_i(t) + \Delta_i(t)^T \bar{B}_i^T P e(t) + e(t)^T P \bar{B}_i \Delta_i(t)$$
$$+ h_i(\hat{x}(t)) \left( e(t)^T \Phi_i^T P e(t) + e(t)^T P \Phi_i e(t) \right) \tag{43}$$

The derivative of the Lyapunov function is composed of quadratic terms in $e(t)$ and of terms crossed in $e(t)$, $\gamma_i(t)$ and $\Delta_i(t)$. In order to express $V(e(t))$ in a quadratic form in $e(t)$. We proceed as follows:

$$\begin{aligned}|\delta_i(t)| &\leq m_i |e(t)| \\ |\Delta_i(t)| &\leq n_i \beta_1 |e(t)| \\ |\psi_k(t)| &\leq \gamma_i |\dot{e}(t)|\end{aligned} \tag{44}$$

By applying the lemma1 we lead to the following inequalities:

$$\begin{aligned}\delta_i(t)^T \bar{A}_i^T P e(t) + e(t)^T P\bar{A}_i \delta_i(t) &\leq \lambda_1 \delta_i(t)^T \delta_i(t) + \lambda_1^{-1} e(t)^T P\bar{A}_i \bar{A}_i^T P e(t) \\ &\leq \lambda_1 m_i^2 e(t)^T e(t) + \lambda_1^{-1} e(t)^T P\bar{A}_i \bar{A}_i^T P e(t)\end{aligned} \tag{45}$$

$$\begin{aligned}\Delta_i(t)^T \bar{B}_i^T P e(t) + e(t)^T P\bar{B}_i \Delta_i(t) &\leq \lambda_2 \Delta_i(t)^T \Delta_i(t) + \lambda_2^{-1} e(t)^T P\bar{B}_i \bar{B}_i^T P e(t) \\ &\leq \lambda_2 n_i^2 \beta_1^2 e(t)^T e(t) + \lambda_2^{-1} e(t)^T P\bar{B}_i \bar{B}_i^T P e(t)\end{aligned} \tag{46}$$

The derivative of the Lyapunov function became:

$$\begin{aligned}\dot{V}(e(t)) = \sum_{i=1}^{r} &\delta_i(t)^T \bar{A}_i^T P e(t) + e(t)^T P\bar{A}_i \delta_i(t) + \Delta_i(t)^T \bar{B}_i^T P e(t) + e(t)^T P\bar{B}_i \Delta_i(t) \\ &+ h_i(\hat{x}(t))\left(e(t)^T \Phi_i^T P e(t) + e(t)^T P\Phi_i e(t)\right)\end{aligned} \tag{47}$$

$$\dot{V}(e(t)) \leq \sum_{i=1}^{r} e(t)^T \begin{pmatrix} h_i(\hat{x}(t))(\Phi_i^T P + P\Phi_i) + (\lambda_1 m_i^2 + \lambda_2 n_i^2 \beta_1^2)I \\ + \lambda_1^{-1} P\bar{A}_i \bar{A}_i^T P + \lambda_2^{-1} P\bar{B}_i \bar{B}_i^T P \end{pmatrix} e(t) \tag{48}$$

The stability of this last equation is assured, for $i = 1,...,r$:

$$e(t)^T \left(h_i(\hat{x}(t))(\Phi_i^T P + P\Phi_i) + (\lambda_1 m_i^2 + \lambda_2 n_i^2 \beta_1^2)I + \lambda_1^{-1} P\bar{A}_i \bar{A}_i^T P + \lambda_2^{-1} P\bar{B}_i \bar{B}_i^T P\right) e(t) < 0 \tag{49}$$

This leads to the following conditions:

$$(A_0 - L_i C)^T P + P(A_0 - L_i C) < -Q \tag{50}$$

$$-Q + (\lambda_1 m_i^2 + \lambda_2 n_i^2 \beta_1^2)I + \lambda_1^{-1} P\bar{A}_i \bar{A}_i^T P + \lambda_2^{-1} P\bar{B}_i \bar{B}_i^T P < 0 \tag{51}$$

Then we suppose variable changement $K_i = PL_i$ and the complement Schur. Then it will be defined as below:

$$A_0^T P + PA_0 - C^T K_i^T - K_i C < -Q \tag{52}$$

$$\begin{bmatrix} -Q + (\lambda_1 m_i^2 + \lambda_2 n_i^2 \beta_1^2)I & P\bar{A}_i & P\bar{B}_i \\ \bar{A}_i^T P & -\lambda_1 I & 0 \\ \bar{B}_i^T P & 0 & -\lambda_2 I \end{bmatrix} < 0 \quad, \lambda_1 > 0 \; et \; \lambda_2 > 0 \tag{53}$$

To have an effective choice we can take the input as a variable to be determined which one will call $\rho$.

By using the complement Schur, inequality (52) will be written:

$$\begin{bmatrix} -Q+\lambda_1 m_i^2 I & P\bar{A}_i & P\bar{B}_i & n_i\lambda_2\rho I \\ \bar{A}_i^T P & -\lambda_1 I & 0 & 0 \\ \bar{B}_i^T P & 0 & -\lambda_2 & 0 \\ n_i\lambda_2\rho I & 0 & 0 & -\lambda_2 I \end{bmatrix} < 0, \ \lambda_1 > 0 \text{ et } \lambda_2 > 0 \tag{54}$$

The presence of the product $\lambda_2\rho$ leads to a nonlinear inequality. To rewrite it in form LMI, we pose $\gamma = \lambda_2\rho$ :

$$A_0^T P + PA_0 - C^T K_i^T - K_i C < -Q \tag{55}$$

$$\begin{bmatrix} -Q+\lambda_1 m_i^2 I & P\bar{A}_i & P\bar{B}_i & n_i\gamma I \\ \bar{A}_i^T P & -\lambda_1 I & 0 & 0 \\ \bar{B}_i^T P & 0 & -\lambda_2 & 0 \\ n_i\gamma I & 0 & 0 & -\lambda_2 I \end{bmatrix} < 0 \tag{56}$$

Knowing $\gamma$ and $\lambda_2$ we can deduce the value $\rho = \dfrac{\gamma}{\lambda_2}$

## 4. DESIGN EXAMPLES

### 4.1. Example 1

This section is dedicated to illustrate the efficiency of the proposed approach. We consider the following example witch illustrate the first theorem:

$$\sum_{k=1}^{2} v_k(z(t)) E_k \dot{x}(t) = \sum_{i=1}^{2} h_i(z(t))(A_i x(t) + B_i u(t))$$
$$y(t) = Cx(t)$$

With $x_0 = \begin{bmatrix} 1 \\ 1 \end{bmatrix}$, $A_1 = \begin{bmatrix} -3 & 1 \\ 1 & -1 \end{bmatrix}$, $A_2 = \begin{bmatrix} -2 & 1 \\ 1 & 0 \end{bmatrix}$, $B_1 = \begin{bmatrix} -2 \\ 1 \end{bmatrix}$, $B_2 = \begin{bmatrix} 1 \\ 1 \end{bmatrix}$, $C = \begin{bmatrix} 1 & 1 \end{bmatrix}$,

$E_1 = \begin{bmatrix} 1 & 0 \\ 0 & 2 \end{bmatrix}$ and $E_2 = \begin{bmatrix} 1 & 0 \\ 0 & 2.01 \end{bmatrix}$.

And the activating functions:

$h_1 = (1-\tanh(x_1))/2$, $h_2 = 1-h_1$ and $v_1 = \cos(x_1)\cos(x_2)$, $v_2 = 1-v_1$.

According to the given procedure, we design the descriptor observer based on theorem 1 via the Matlab LMI toolbox. Then we obtain $P_1$ : and $P_3$ :

$$P_1 = \begin{bmatrix} 0.9875 & -0.0266 \\ -0.0266 & 1.2489 \end{bmatrix}, \ P_3 = \begin{bmatrix} 0.2561 & -0.0373 \\ 0.0394 & 0.3161 \end{bmatrix}.$$

The corresponding observer gain matrices are:

$$L_1 = \begin{bmatrix} -0.4898 \\ 0.4323 \end{bmatrix}, \ L_2 = \begin{bmatrix} 0.5255 \\ -0.1566 \end{bmatrix}.$$

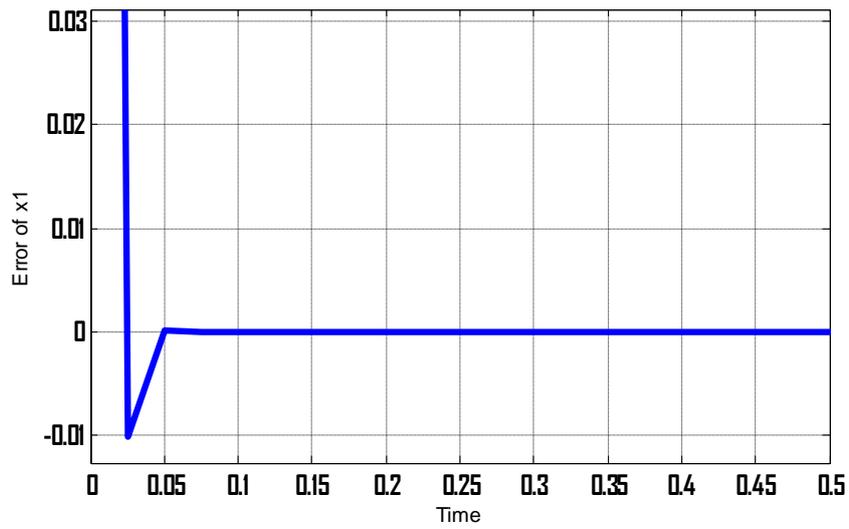

Figure1. Error estimation for x1

We conclude from the simulation results that the observe design shows the essential aims of this method. The estimation error of $x_1 - \hat{x}_1$ is shown in Figure 1 and the error of $x_2 - \hat{x}_2$ in Figure 2 are close to zero quickly.

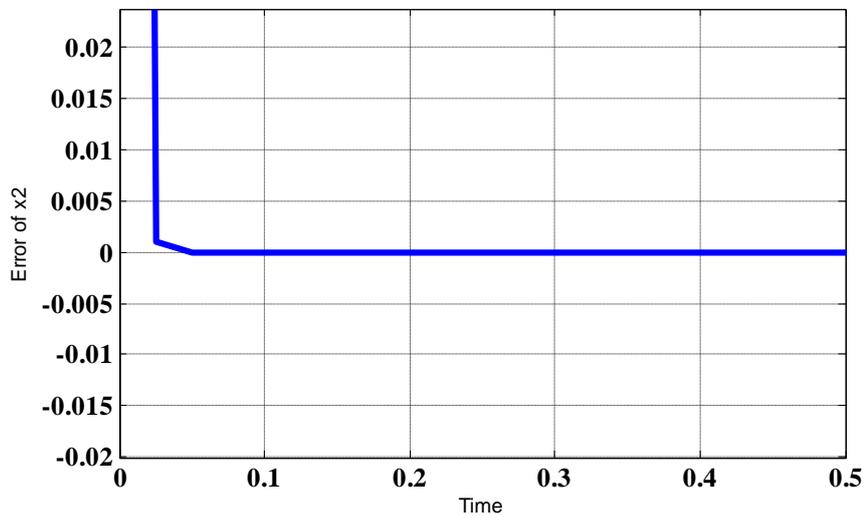

Figure2. Error estimation for x2

## 4.2. Example 2

The theorem 2 is illustrated by this example:

$$E\dot{x}(t) = \sum_{i=1}^{3} h_i(z(t))(A_i x(t) + B_i u(t))$$
$$y(t) = Cx(t)$$

With $x_0 = \begin{bmatrix} 1 \\ 1 \\ 1 \end{bmatrix}$, $A_1 = \begin{bmatrix} -2 & 1 & 1 \\ 1 & -3 & 0 \\ 2 & 1 & -6 \end{bmatrix}$, $A_2 = \begin{bmatrix} -3 & 2 & -2 \\ 5 & -3 & 0 \\ 0.5 & 0.5 & -4 \end{bmatrix}$, $B_1 = \begin{bmatrix} 1 \\ 0.5 \\ 0.5 \end{bmatrix}$, $B_2 = \begin{bmatrix} 0.5 \\ 1 \\ 0.5 \end{bmatrix}$, $C = \begin{bmatrix} 1 & 1 & 1 \\ 1 & 0 & 1 \end{bmatrix}$

and

$$E = \begin{bmatrix} 1 & 2 & 0 \\ 0 & 2 & 0 \\ 0 & 0 & 1 \end{bmatrix}.$$

And the activating functions:

$$h_1 = (1 - \tanh(x_1))/2, \quad h_2 = 1 - h_1.$$

According to the given procedure, we design the descriptor observer based on theorem 2 via the Matlab LMI toolbox. Then we obtain:

$$P = \begin{bmatrix} 1.2594 & 0.2698 & 0.4471 \\ 0.2698 & 1.4447 & -0.0991 \\ 0.4471 & -0.0991 & 0.7367 \end{bmatrix} \text{ and } Q = \begin{bmatrix} 1.5534 & -0.0000 & -0.0300 \\ -0.0000 & 1.5235 & -0.0000 \\ -0.0300 & -0.0000 & 1.5534 \end{bmatrix}$$

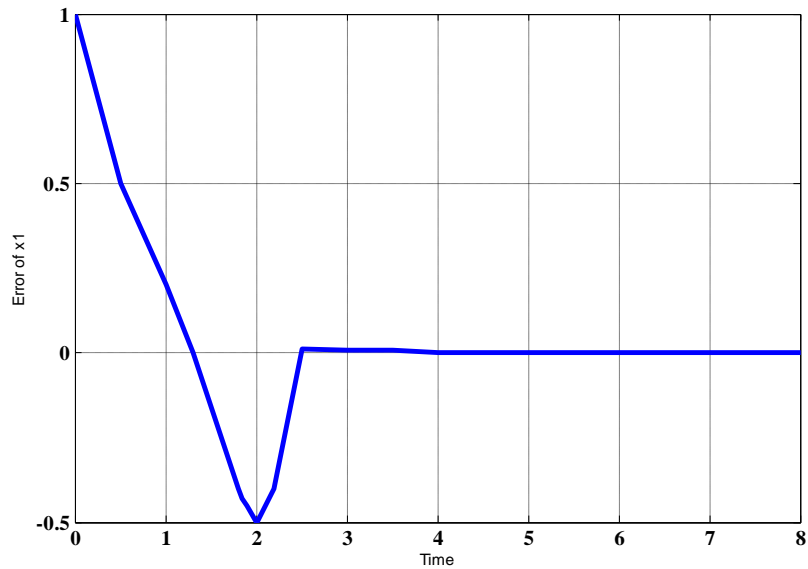

Figure3. Error estimation for x1

The corresponding observer gain matrices are:

$$L_1 = \begin{bmatrix} 118.4894 & 0.3110 \\ -101.6108 & 336.9255 \\ -608.0136 & 908.4754 \end{bmatrix} \text{ and } L_2 = \begin{bmatrix} -421.6323 & 106.0784 \\ 93.6932 & 266.3481 \\ -273.6247 & 165.1607 \end{bmatrix}$$

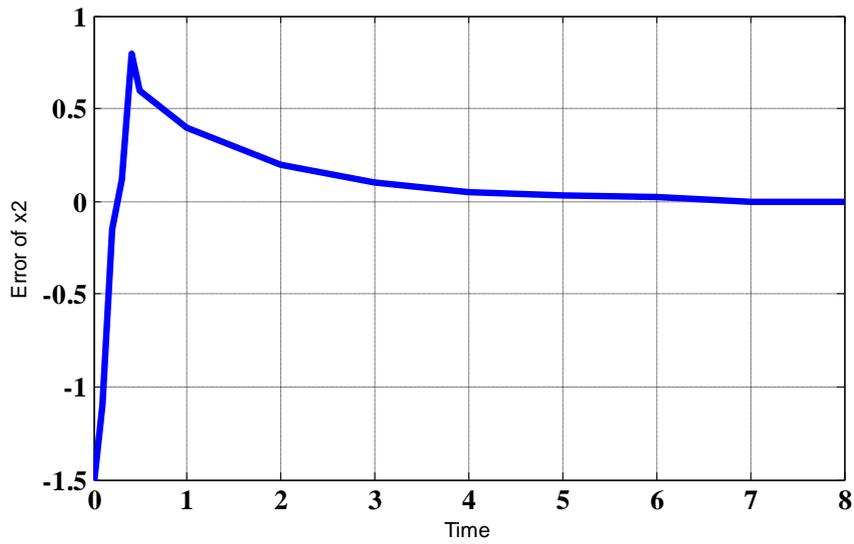

Figure4. Error estimation for x2

The scalars are of values: $\lambda_1 = 0.5863$, $\lambda_2 = 0.0094$ and $\gamma = 0.1575$.

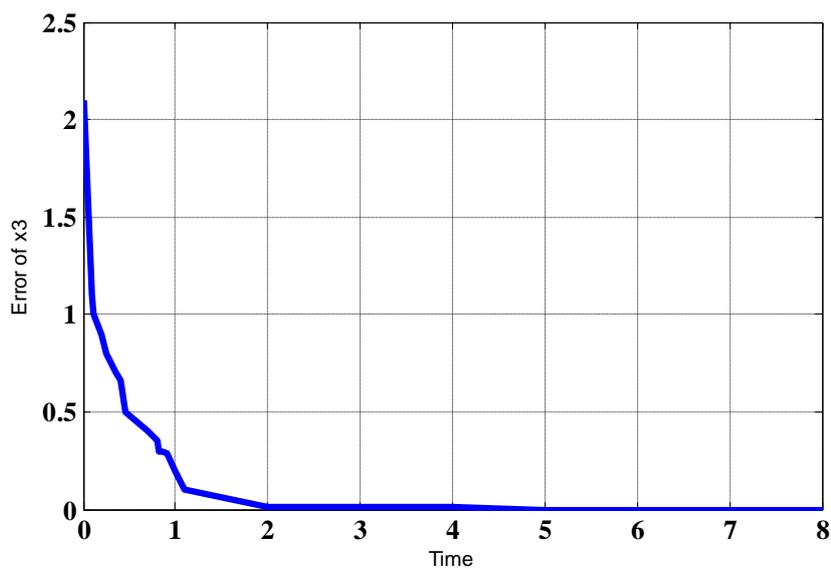

Figure5. Error estimation for x3

The results of simulation corresponding to the evolution of the state estimation error are presented in Figure 3, Figure 4 and Figure 5.

## 5. CONCLUSIONS

The design descriptor observer was studied in this paper. The considered systems are modeled in the Takagi-Sugeno descriptor structure with measurable and unmeasurable premise variables. The strategy is based on the use of the Lipschitz condition. The stability is studied with the Lyapunov theory and a quadratic function that allows to derive conditions ensuring the convergence of the state estimation error. The existence conditions are expressed in terms of LMIs that can be solved with LMIS Toolbox in Matlab.